\documentclass[conference]{IEEEtran}

\IEEEoverridecommandlockouts

\usepackage{cite}
\usepackage{amsmath,amssymb,amsfonts}
\usepackage{graphicx}
\usepackage{textcomp}
\usepackage{xcolor}
\def\BibTeX{{\rm B\kern-.05em{\sc i\kern-.025em b}\kern-.08em
    T\kern-.1667em\lower.7ex\hbox{E}\kern-.125emX}}

\usepackage{booktabs} 
\usepackage{tabularx} 
\usepackage{caption} 
\usepackage{subcaption}

\usepackage{algorithm}
\usepackage{algpseudocode}
\usepackage{amsmath}
\usepackage{comment}

\usepackage{pifont}
\newcommand{\cmark}{\ding{51}}%
\newcommand{\xmark}{\ding{55}}%

\begin{document}

\title{Should we use model-free or model-based control? A case study of battery management systems\\
\thanks{Mohamad Fares El Hajj Chehade and Young-ho Cho contribute equally to this paper. (Corresponding author: Mohamad Fares El Hajj Chehade.)}
}

\author{\IEEEauthorblockN{Mohamad Fares El Hajj Chehade\textsuperscript{*}, Young-ho Cho\textsuperscript{*}, Sandeep Chinchali, and Hao Zhu}
\IEEEauthorblockA{Chandra Department of Electrical and Computer Engineering \\
The University of Texas at Austin\\
Austin, TX, USA \\ 
\{chehade, jacobcho, sandeepc, haozhu\}@utexas.edu}
}

\maketitle



\begin{abstract}
    Reinforcement learning (RL) and model predictive control (MPC) each offer distinct advantages and limitations when applied to control problems in power and energy systems.
    Despite various studies on these methods, benchmarks remain lacking and the preference for RL over traditional controls is not well understood.
    In this work, we put forth a comparative analysis using RL- and MPC-based controllers for optimizing a battery management system (BMS). The BMS problem aims to minimize costs while adhering to operational limits. by adjusting the battery (dis)charging  in response to fluctuating electricity prices over a time horizon.
    %
    The MPC controller uses a learning-based forecast of future demand and price changes to formulate a multi-period linear program, that can be solved using off-the-shelf solvers. 
    Meanwhile, the RL controller requires no time-series modeling but instead is trained from the sample trajectories using the proximal policy optimization (PPO) algorithm.
    Numerical tests compare these controllers across optimality, training time, testing time, and robustness, providing a comprehensive evaluation of their efficacy.
    RL not only yields optimal solutions quickly but also ensures robustness to shifts in customer behavior, such as changes in demand distribution. However, as expected, training the RL agent is more time-consuming than MPC.
\end{abstract}

\begin{IEEEkeywords}
Battery management system, reinforcement learning, model predictive control, time-series forecast. 
\end{IEEEkeywords}


\section{Introduction}
\label{sec:introduction}

In the domain of control problems, reinforcement learning (RL) and model-predictive control (MPC) have their merits and drawbacks. RL deals with agents in uncertain settings and optimizes the actions over infinite horizons without modeling the dynamics of the environment. It excels in environments with large modeling errors but suffers from data inefficiency and safety due to possibly undesirable actions. On the other hand, MPC optimizes actions over a finite horizon, with heavy reliance on accurate environmental models. It is less concerned with safety and data efficiency, due to the deterministic nature of the optimization problem and its use of hard constraints. However, it could perform poorly due to modeling inaccuracies and the limitations of a finite horizon\cite{BMS, Gaussian, cruise_control}.

In power and energy systems, both RL and MPC have been explored for the battery control problem, because of its sequential decision-making nature. In the problem of battery management system (BMS), the charging and discharging schedule of an energy storage system is controlled to minimize the cost of electricity purchase over a certain period. For example, one work based on MPC uses a long short-term memory (LSTM) forecast to model the system; then, the multi-period optimization problem is solved \cite{MPC_example}. Despite showing its superiority over conventional methods like dynamic programming and fuzzy logic, this work does not check the impact of forecasting errors on the optimality of MPC and other methods. On the other hand, RL in \cite{RL_example}, particularly deep-Q networks (DQN), has improved the battery's operation and minimized its degradation. However, there is a lack of benchmark in general for justifying the choice of RL over currently used control methods. 

In this work, a comparative analysis is conducted between the model-based MPC and the model-free RL approaches to the battery control problem. The closest to our work is \cite{Zamzam}, which compares the performance of DQN with one-step MPC. Nonetheless, the comparisons therein are limited in three aspects. First, DQNs have been outperformed by several policy-based algorithms, such as proximal policy optimization (PPO) and deep deterministic policy gradient (DDPG), which are currently the popular RL algorithms in power and energy systems \cite{policy_based_example1, policy_based_example2}. Second, one-step MPC is essentially a greedy algorithm that does not benefit from planning over multiple time steps, thereby compromising the optimality of the MPC results. Third, the comparison is solely based on the control cost, and does not include some important criteria, such as robustness and computational efficiency. To the best of our knowledge, our work is the first to address these three aspects and conduct a full comparative analysis between RL and MPC for BMS. The main contributions of this work are summarized below:

\begin{enumerate}
    \item We study the optimality of the decisions made by RL and MPC for the battery control problem. We adopt the well-known, widely used algorithms for each approach. 
    \item We extend the basis of the comparison from control cost to include multiple other criteria: data efficiency, online computation time, and robustness. 
    \item Our results show that RL performs better in terms of optimality and online computation time. It is also more robust to shifts in the demand distribution. However, it is much more data-inefficient than MPC, whose optimality is drastically affected by forecasting errors. 
    \item Our methodology and analysis go beyond the BMS problem and could be applicable to general control problems in power and energy systems. 
\end{enumerate}

The rest of the paper is organized as follows: Section \ref{sec:problem_formulation} defines the problem, Sections \ref{sec:MPC} and \ref{sec:RL} respectively introduce the MPC and RL approaches, Section \ref{sec:simulations} discusses the main results, and Section \ref{sec:conclusion} concludes the paper.  

\section{Preliminaries}
\label{sec:problem_formulation}

\begin{figure}[t!]
   \centering
   \includegraphics[width=0.8\linewidth]{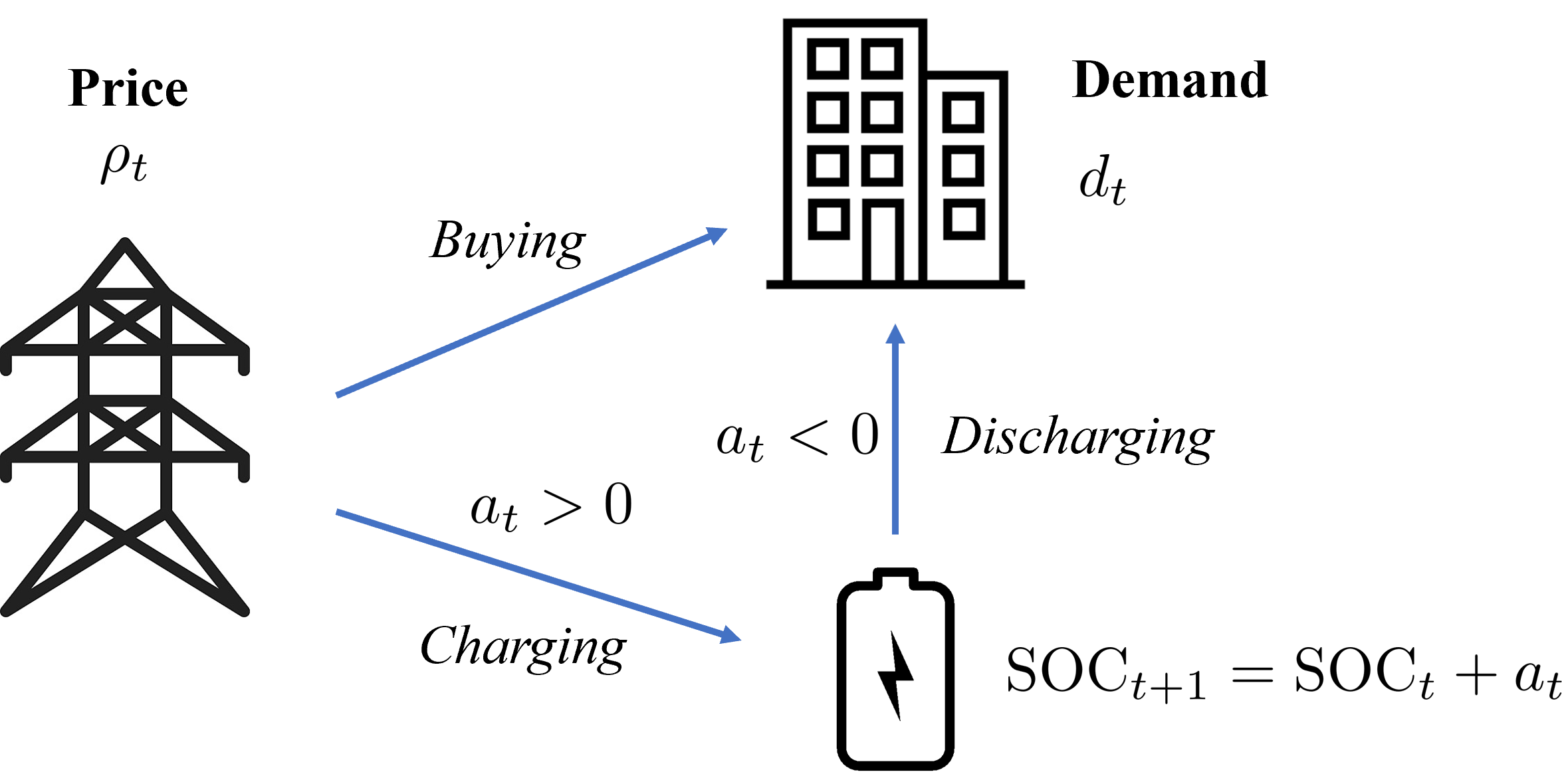}
   \caption{The components of the system. At a given time $t$, the load has a demand $d_t$ and the utility sells electricity at a price $\rho_t$. To minimize the total cost of purchasing electricity over a certain period, the battery controller chooses between charging from the grid ($a_t > 0$) and discharging to the load ($a_t <0$).} 
   \label{fig:bms_diagram}
\end{figure}

The battery control problem is formulated for a system of three components: a load with a time-varying demand for electricity, denoted by $d_t$ (in kW) for a given time $t$, an electric utility that sells electricity at a time-varying price $\rho_t$ (in \$/kWh), and a battery storage system of a capacity $E$ (in kWh) and state of charge $\text{SOC}_t$. The latter is capable of charging electricity from the utility and discharging to the load at a rate $a_t = a_\text{max}$ (in kW), as long as $\text{SOC}_t$ remains within operational limits $\text{SOC}_\text{min}$ and $\text{SOC}_\text{max}$. The electricity tariff $g_t$ at a given time $t$ is then linear in $a_t$: $g_t = \rho_t(d_t + a_t)$. The interaction between the three components of the system is best illustrated in Fig. \ref{fig:bms_diagram}. 

To minimize the cost of purchasing electricity, the battery must be discharged during periods of high tariffs, and charged otherwise. However, due to the battery's limited capacity, the controller must plan over a certain time horizon, to see when charging and discharging are most economical. This results in a multi-step optimization problem with cost function $f$, where $f$ is the sum of individual tariffs at the considered timesteps: $f = \sum_{t} g_t$. Despite the linearity of the cost function in terms of the decision variables $a_t$, the problem is non-trivial due to the uncertainties in the demand $d_t$ and prices $\rho_t$. As a result, two approaches are considered for solving the battery control problem: model-based and model-free.

\section{Model-Predictive Control (MPC)}
\label{sec:MPC}
The MPC problem aims at minimizing the operating cost of the energy system over a finite time horizon $t \in \mathcal{T}=\{1,2, \ldots, T\}$, while maintaining the battery within its operational limits. At each time step $t \in \mathcal{T}$, the first element $b_{t|t}$ in the vector of solutions obtained from each optimization problem is stored. Subsequently, another optimization problem is solved starting at the time step $t+1$. The demand and price in the problem are sampled from two time-series distributions: $\rho_{{t+k}|t} \sim f_\rho(t+k, \theta)$ and $d_{{t+k}|t} \sim f_d(t+k, \mu)$ for all $k \in \{1,2,3,\ldots,T\}$, where $\theta$ and $\mu$ are the parameters of the price and demand distributions, respectively. We obtain estimated values for price and demand from forecasters, i.e., $\hat{\rho}_{t+k}|t \sim f_{\hat{\rho}}(t+k, \hat{\theta})$ and $\hat{d}_{t+k}|t \sim f_{\hat{d}}(t+k, \hat{\mu})$ for all $t \in \mathcal{T}$, where $\mathcal{T}$ is the full horizon of the problem.

To tackle the inherent unpredictability of future demand and electricity prices, we employ a predictive framework relying on an LSTM network, well-suited for sophisticated time series forecasting. This neural network acts as a forecaster, projecting demand and cost factors from the current time step $t+1$ to a specified horizon $\mathcal{T}$. The MPC framework leverages the forecasted data to make informed decisions about battery charging and discharging, aiming to minimize electricity expenses and stabilize battery energy levels. To enhance the reliability of these forecasts, we engage in iterative refinement of the predictive models. This critical step involves continuously aligning the models with observed data, thereby reducing forecasting errors and improving decision-making accuracy in the MPC framework.

Given the predicted electricity cost $\hat{\rho}_{t+k|t}$ at time $t+k$, which is predicted at time $t$, the BMS optimization problem seeks to minimize the total cost while satisfying the operational limits, as given by:
\begin{subequations}
\label{eq:mpc problem}
\begin{align}
    & \underset{a_{t|t}, \ldots, a_{t+T|t}}{\text{min}}
    & & \sum_{k=0}^{T} \hat{\rho}_{t+k|t} (E \cdot a_{t+k|t} + \hat{d}_{t+k|t}) \\
    & \text{subject to}
    & & \text{SOC}_{t+k+1|t} = \text{SOC}_{t+k|t} + a_{t+k|t},\label{mpc1}\\
    &&& \text{SOC}_{\min} \leq \text{SOC}_{t+k+1|t} \leq \text{SOC}_{\max},\label{mpc2}\\
    &&& \|a_{t+k|t}\| \leq a_{\text{max}},\label{mpc3}\\
    &&& \forall k \in \{0, \ldots, T\},~\forall t \in \mathcal{T}.
\end{align}
\end{subequations}
The constraints in \eqref{mpc1} represent the remaining battery level, $\text{SOC}_{t+k+1|t}$, at $t+k+1$ after making a charging or discharging decision at $t+k$. The battery's operational limits in \eqref{mpc2} range from $\text{SOC}_{\min}$ to $\text{SOC}_{\max}$. Moreover, the maximum allowable charge/discharge rate is limited by the constant $a_{\text{max}}$ (in kW), as denoted in \eqref{mpc3}. The problem \eqref{eq:mpc problem} is a multi-step linear optimization that can be efficiently solved by off-the-shelf solvers.

\section{Reinforcement Learning (RL)}
\label{sec:RL}




Due to its sequential decision-making nature and Markovian state space, the problem can be modeled as a Markov decision process (MDP) \cite{Puterman}. An MDP is defined as the tuple $(\mathcal{S}, \mathcal{A}, p, r, \gamma)$, where $\mathcal{S}$ and $\mathcal{A}$ are sets of states and actions, respectively. For an action $a \in \mathcal{A}$ that an agent takes in a given state $s \in \mathcal{S}$, $p(.|s,a)$ models the distribution over the next states. For the transition $s \xrightarrow{a} s'$, the agent receives a scalar reward $r(s,a,s')$. Finally, the discount factor $\gamma \in [0,1]$ can be used to give smaller weights for future rewards. The elements of the MDP tuple for the BMS problem are defined below: 

\begin{itemize}
    \item \textbf{State space} (\(\mathcal{S}\)): The state is characterized by the tuple \((\text{SOC}_t, \rho_t, d_t, h_t, D_t)\), where:
    \begin{itemize}
        \item \(\text{SOC}_t\) represents the state of charge of the battery at time \(t\).
        \item \(\rho_t\) is the electricity price at time \(t\).
        \item \(d_t\) indicates the energy demand at time \(t\).
        \item \(h_t\) is the hour $t$ of the day.
        \item \(D_t\) distinguishes between weekdays and weekends.
    \end{itemize}
    
    \item \textbf{Action space} (\(\mathcal{A}\)): The action space is one-dimensional and represents the control variable $a_t$, defined as the amount of charge/discharge at a given time $t$. Since the optimal solution is always one of three values: charge with the maximum amount ($a_t$ = $a_\text{max}$ = 0.1E), discharge with the maximum amount ($a_t$ = - $a_\text{max}$ = - 0.1E), or idle ($a_t$ = 0), the action space is discrete $A = \{-0.1E, 0, 0.1E\}$.
    
    \item \textbf{Transition function} (\(p\)): Time elements in the state space incur deterministic transitions. The state of charge progression is described below:
    \begin{align}
        \text{SOC}_{t+1} = \begin{cases} 
            \text{SOC}_{\min} & \text{if } \text{SOC}_t + a_t < \text{SOC}_{\min} \\
            \text{SOC}_{\max} & \text{if } \text{SOC}_t + a_t > \text{SOC}_{\max} \\
            \text{SOC}_t + a_t & \text{otherwise} 
        \end{cases}
    \end{align}
This ensures that SOC stays within the operational limits \(\text{SOC}_{\min}\) and \(\text{SOC}_{\max}\). On the other hand, the electricity price and energy demand are assumed to be drawn from two time series distributions: $\rho_t \sim f_\rho$ and $d_t \sim f_d$. Modeling $f_\rho$ and $f_d$ is challenging and may incur significant errors. Fortunately, model-free RL allows for solving the problem without any knowledge of the time-series distributions.  
\item \textbf{Reward function} (\(r\)): the reward is defined as the negative of the instantaneous electricity cost of the energy consumed. The power consumed is used to satisfy the load demand and to charge the battery: $r_t = - \rho_t (d_t + a_t E)$.

\item \textbf{Discount factor} (\(\gamma\)): Since electricity costs at future time steps matter as much as costs in the present, no discounting is applied, i.e. $\gamma = 1$. 
\end{itemize}

In an MDP framework, an RL agent takes its actions based on a policy distribution $\pi$, i.e. $a_t \sim \pi(\cdot|s_t)$, where $\pi(a|s)$ is the probability of choosing action $a$ from state $s$ under policy $\pi$. Given that the agent is at state $s_t$ at a given time $t$, the return $G_t$ is defined as the total reward the agent receives starting at time $t$ and until the end of the trajectory, when taking actions based on policy $\pi$. The expected value of the return is known as the value function $V^{\pi}(s_t)$. The goal of the RL problem is to find the optimal policy $\pi^*$, which maximizes the expected total reward the agent receives over its trajectory, i.e. $\pi^* = \max_\pi V^{\pi}(s) \: \forall s \in \mathcal{S}$.



\subsection{Methodology}
\begin{figure}[t!]
   \centering
   \includegraphics[scale=0.30]{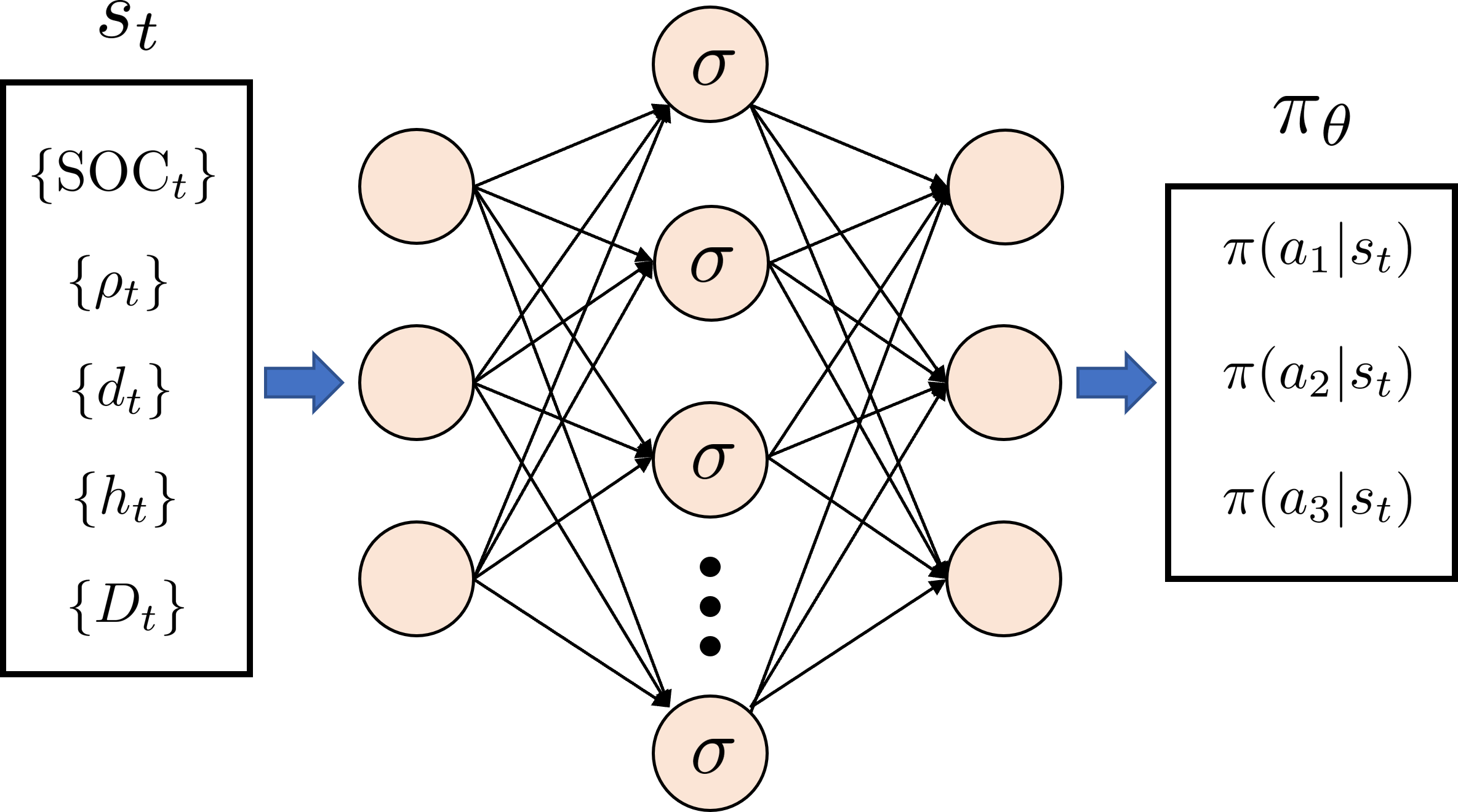}
   \caption{The policy network $\pi_{\theta}$. The state vector $s_t$ propagates through the neural network layers to output the probability masses $\pi_{\theta}(a|s_t) \: \forall a \in \mathcal{A}$. $\sigma$ is a non-linear activation function to allow the network to represent non-linear input-output relationships. A softmax function is applied at the output to convert the raw numbers (logits) to probabilities.}
   \label{fig:policy_network}
\end{figure}

In modern RL, policy-based deep RL methods have enjoyed more success than classical value-based methods \cite{TRPO, A3C, SAC}. In the discrete action-space setting, the policy distribution is represented by a deep neural network with parameters $\theta$, as shown in Fig. \ref{fig:policy_network}. The input vector to the network is the state $s$, and the output vector consists of the probability masses $\pi_{\theta}(a|s)$. For a given state $s_t$, the action is sampled from the output vector distribution: $a_t \sim \pi_{\theta}(\cdot|s_t)$. The optimal policy $\pi^* = \pi_{\theta^*}$ is obtained by maximizing the expected return received along its experienced trajectories. In other words, if the agent experiences a trajectory $\tau$, which produces a large return $G^{\tau}_0$, the actions that yield such a trajectory are enforced by the policy network, i.e. their probabilities increase. The objective function is formally represented below:

\begin{align}
\label{eq:objective}
\max_{\theta} J(\theta) &= \max_{\theta} \mathbb{E}_{\tau} [G_0 \mid s_0] \\
&= \max_{\theta} \sum_\tau P(\tau) G_0
\end{align}

where $P(\tau)$ is the probability of trajectory $\tau$, and is a function of the actions chosen by the agent, i.e. $\pi_{\theta}$ and the environment transitions $p$.

This can be achieved with gradient ascent in the $\theta$ space:

\begin{align}
\label{eq:gradient_ascent}
\theta &\leftarrow \theta + \alpha \sum_{t=0}^T \nabla_{\theta} \log \pi_{\theta}(a_t \mid s_t) G_t
\end{align}

where $\alpha$ is the step size, and $T$ is the number of steps taken in the trajectory. 

To reduce the variance in the return during learning, an action-independent baseline is subtracted from the return in \eqref{eq:gradient_ascent}. In most settings, this baseline is the value function $V(s_t)$, and the objective function becomes:

\begin{align}
\label{eq:gradient_ascent_with_baseline}
\theta &\leftarrow \theta + \alpha \sum_{t=0}^T \nabla_{\theta} \log \pi_{\theta}(a_t \mid s_t) \left(G_t - V(s_t)\right)
\end{align}

The value function itself can be represented as a separate neural network with parameter $\phi$, and is optimized by minimizing the temporal difference (TD) error:

\begin{align}
\label{eq:bootsrapping}
\min_{\phi} \left( r + \gamma V_{\phi}(s') - V_{\phi}(s) \right)^2
\end{align}

In some situations, large parameter updates result in unstable learning and potential divergence. As a remedy, these updates can be ensured to be within a trust region, by enforcing the KL-divergence between the parameters before and after the update, $D_{KL}(\theta_{new}||\theta_{old})$, to be within a threshold $\delta$. Therefore, the parameters take the largest step (towards the steepest ascent) that keeps them inside trust region. This translates to the constrained optimization problem below:

\begin{equation}
\label{eq:trust_region_problem}
\begin{aligned}
\theta_{\text{new}} = \arg \max_{\theta} & \quad J(\theta) \\
\text{s.t.} & \quad D_{\text{KL}}(\theta \parallel \theta_{\text{old}}) \leq \delta
\end{aligned}
\end{equation}

An approximate form of this problem is solved by the proximal policy optimization (PPO) algorithm \cite{PPO}, whose pseudocode is shown in Algorithm \ref{alg:PPO}. 

\begin{algorithm}{!t}
\caption{Proximal Policy Optimization (PPO)}
\label{alg:PPO}
\begin{algorithmic}[1]

\State Initialize parameters $\theta$ for policy $\pi_\theta$ and $\phi$ for value function $V_\phi$
\Repeat
    \State Collect trajectories $D_k = \{s_t, a_t, r_t, s'_t\}$ using $\pi_\theta$
    \State Compute $G_t$, $A_t = G_t - V_\phi(s_t)$
    \For{each epoch over $D_k$}
        \State Update $\theta$ by (approximately) solving \eqref{eq:trust_region_problem}
        \State Update $\phi$ minimizing $\left(r_t + \gamma V_\phi(s'_t) - V_\phi(s_t)\right)^2$
    \EndFor
\Until{convergence}
\end{algorithmic}
\end{algorithm}

\section{Numerical Comparisons}
\label{sec:simulations}



In this section, we compare the performance of reinforcement learning (RL) and model-predictive control (MPC) on a residential building. We consider two stages in the evaluation: training and testing. In the training phase, the RL agent learns a policy network, as outlined in Algorithm \ref{alg:PPO}, while a forecaster for price and demand is trained for MPC. In the testing phase, RL and MPC are run on a new data set. The performance of each method is judged based on four criteria:

\begin{enumerate}

\item \textbf{control performance (optimality):} the total cost incurred on the test dataset.
\item \textbf{data efficiency:} the data needed to train the RL agent or forecaster 
\item \textbf{testing time:} the time needed to make control decisions for the test dataset
\item \textbf{robustness:} the degree of optimality when the distribution of the model changes from the training dataset to the testing dataset

\end{enumerate}

For this purpose, two datasets (each with training and testing sets) are considered. In the first dataset, the model of the system remains the same over the training and testing sets, and the first three criteria are tested. In the second dataset, the distribution of the demand changes from the training to the testing set, therefore allowing us to examine the robustness of each controller.

As a baseline, we consider a pre-defined policy that discharges the battery when the price is above average (computed based on the training dataset) and charges the battery otherwise - as long as the state of charge is within operational limits. 

In both experiments, the data is hourly, and the horizon for MPC is 24 hours. An LSTM network is trained in Tensorflow as the forecaster, while StableBaselines3 \cite{SB3} with default parameters is used to train the RL agent. Five runs are performed for each method, for which a 95\% confidence interval of the cost is computed. The simulations took place on a machine equipped with an Intel\textsuperscript{\textregistered} CPU @ 2.10 GHz and 32 GB RAM.


\begin{table}[tbp!]
    \centering
    \caption{The performance metrics for the different controllers. The ground truth is computed by considering the actual values of demand and prices and optimizing over the entire test set. RL performs best in terms of optimality and testing time, while MPC is more data-efficient. The optimality of MPC (with true values of demand and price) indicates that errors in the regular MPC model are due to the forecaster.}
    \label{tab:original_dataset}
    \begin{subtable}{\columnwidth}
        \centering
        \caption{Optimality (control cost)}
        \begin{tabular}{lcc}
            \toprule
            \textbf{Controller} & \textbf{Cost (\$)} & \textbf{Optimality Gap} \\
            \midrule
            \textbf{RL} & $\textbf{85,100} \mathbf{\pm} \textbf{200}$ & $\textbf{7.36\%}$ \\
            MPC & $89,650 \pm 665 $ & $13.10\%$ \\
            MPC (exact model) & $79,268$ & $0\%$ \\
            Baseline & $90,989$ & $14.77\%$ \\
            Ground truth & $79,268$ & - \\
            No BMS & $114,065$ & $43.89\%$ \\
            \bottomrule
        \end{tabular}
    \end{subtable}

    \vspace{1em} 

    \begin{subtable}{\columnwidth}
        \centering
        \caption{Data Used and Testing Time}
        \begin{tabular}{lcc}
            \toprule
            \textbf{Controller} & \textbf{Data Used} & \textbf{Testing Time (s)} \\
            \midrule
            RL & $3 \times 10^6$ & $5$ \\
            MPC & $3 \times 10^3$ & $30$ \\
            \bottomrule
        \end{tabular}
    \end{subtable}
\end{table}

\subsection{Dataset 1: Consistent Model through Training and Testing Sets}
The load corresponds to a residential building,  and the real-time (RT) price are from the Pennsylvania, New Jersey, and Maryland (PJM) region. The training and testing sets are for July 2017 and 2018 respectively.

Table \ref{tab:original_dataset} shows the performance of each controller based on the first three criteria. While the table shows that any control method that employs a battery storage system is economic (as compared with no BMS), RL achieves the highest degree of optimality, as compared with MPC and the baseline. This reinforces the capabilities of trust-region policy-based methods in reaching the optimal solution without any model of the environment \cite{PPO, TRPO}. On the other hand, MPC suffers from modeling errors incurred by the forecaster and performs just better than the baseline pre-defined policy. The errors in the forecast are verified by the optimality of the MPC solution when considering the actual values for demand and price, rather than the forecasted ones. 

In the meantime, RL requires about 1000 times more data than MPC, as the RL agent requires a lot of interactions with the environment during training \cite{Gaussian}. Nonetheless, training is performed only once, after which RL is 6 times faster during the testing phase. This is because decision-making for RL involves a forward propagation over the policy network (Fig. \ref{fig:policy_network}), while MPC has to solve a multi-step optimization problem every time a decision is made. 

            



\subsection{Dataset 2: Distributional Shifts in the Demand from the Training to the Testing Sets}

\begin{figure}[t!]
   \centering
   \includegraphics[scale=0.30]{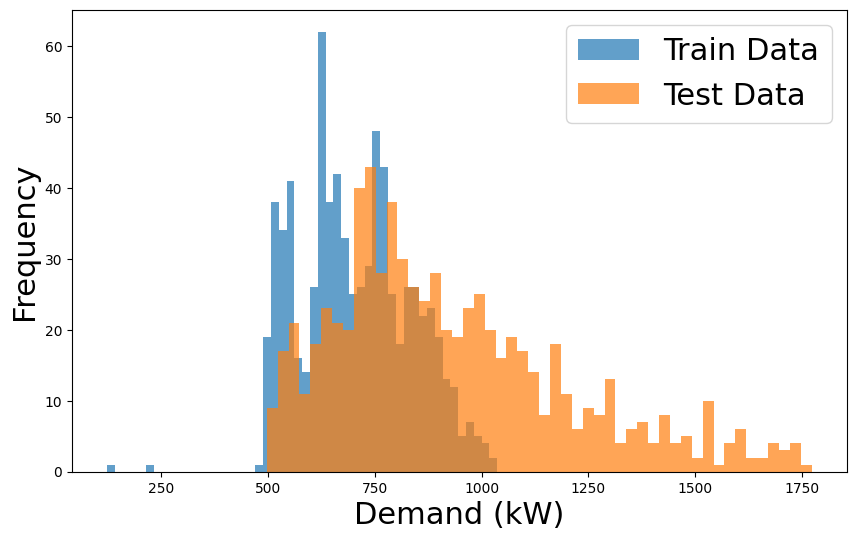}
   \caption{The shift in the distribution of the demand from the training to the testing datasets.}
   \label{fig:distrib_shift}
\end{figure}

The training and testing data are for July and August of 2017, respectively. The price data is again retrieved from PJM, while the demand is for a commercial building in South Korea. Fig. \ref{fig:distrib_shift} illustrates the differences in the distributions of demand between training and testing. 

Table \ref{tab:robustness} displays the total cost incurred by each controller, where RL is shown to maintain its high level of optimality. This not only emphasizes the advantages of model-free learning utilized by RL, but also RL's ability to learn general rules that transcend changes in the model. On the other hand, the errors in the forecaster are magnified by this distributional shift, which resulted in a relatively high control cost for MPC.

\begin{table}[tbp!]
    \centering
    \caption{The optimality of each controller when considering distributional shifts in the demand from training to testing. RL is robust to such shifts, as it maintains a high level of optimality due to its generalization capabilities. The performance of MPC is again drastically affected by forecasting errors.}
    \label{tab:robustness}
    \begin{tabular}{lccc}
        \toprule
        \textbf{Controller} & \textbf{Cost (\$)} & \textbf{Optimality Gap} \\
        \midrule
        \textbf{RL} & $\textbf{158,977} \mathbf{\pm} \textbf{389}$ & $\textbf{2.58\%}$ \\
        MPC & $161,806 \pm 740 $ & $4.40\%$ \\
        MPC (exact model) & $154,981$ & $0\%$ \\
        Baseline & $166,490$ & $7.43\%$ \\
        Ground Truth & $154,981$ & - \\
        No BMS & $179,958$ & $16.12\%$ \\
        \bottomrule
    \end{tabular}
\end{table}

\section{Conclusion}
\label{sec:conclusion}


To sum up, we have developed and compared the model-based MPC and model-free RL controllers for the battery management system.
Although RL requires longer training time compared to MPC, it can achieve optimal control and offers faster testing time than MPC.
We summarize the general comparison of RL and MPC for optimal control in Table~\ref{conc}.
\begin{table}[tbp!]
    \centering
    \caption{Comparison of RL and MPC for the optimal control of the battery storage system. RL yields more optimal solutions in less time. Furthermore, its solutions are robust to changes in customer behavior, represented by shifts in the demand distribution. On the other hand, training the RL agent requires much more time than MPC.}
    \label{conc}
    \begin{tabular}{lccc}
        \toprule
        \textbf{Comparison criteria} & \textbf{RL} & \textbf{MPC} \\
        \midrule
        Optimality & \cmark & \xmark \\
        Data Efficiency & \xmark & \cmark \\
        Online Computation & \cmark & \xmark \\
        Robustness & \cmark & \xmark \\
        \bottomrule
    \end{tabular}
\end{table}
Our future research directions include considering more generalized energy management systems with photovoltaic panels and multiple loads, as well as implementing different types of forecasters to handle uncertainties arising from those components.



\bibliographystyle{IEEEtran}
\bibliography{ref} 

\end{document}